\documentclass[english,aps,pra,showpacs,twocolumn,superscriptaddress,floatfix]{revtex4}

\usepackage{amsmath}
\usepackage{amssymb}
\usepackage{amsfonts}
\usepackage{bbm}
\usepackage{epsfig}
\usepackage{grffile}	
\usepackage{babel}

\usepackage{color}
\usepackage[draft]{hyperref}

\newcommand{\ud}{\mathrm{d}}
\newcommand{\id}{\mathbbm{1}}
\newcommand{\Tr}{\operatorname{Tr}}
\newcommand{\Span}{\operatorname{span}}
\newcommand{\bra}{\langle}
\newcommand{\ket}{\rangle}
\newcommand{\mc}[1]{\mathcal{#1}}
\newcommand{\pdag}{{\phantom{\dag}}}

\renewcommand{\H}{\mc{H}}
\newcommand{\hH}{\hat{H}}
\newcommand{\hb}{\hat{b}}
\newcommand{\hc}{\hat{c}}
\newcommand{\hA}{\hat{A}}
\newcommand{\hB}{\hat{B}}
\newcommand{\dmp}{\rho}
\newcommand{\dm}{\hat\rho}
\newcommand{\aux}{{\operatorname{aux}}}
\newcommand{\LP} {{\operatorname{LP}}}
\renewcommand{\max}{{\operatorname{max}}}
\renewcommand{\vec}[1]{{\boldsymbol{#1}}}

\newcommand{\normS}[1]{\Vert #1\Vert}

\definecolor{grey}{rgb}{.5,.5,.5}
\definecolor{dred}{rgb}{.9,0,0}

\newcommand{\lmu} {Department of Physics and Arnold Sommerfeld Center for Theoretical Physics,
Ludwig-Maximilians-Universit{\"a}t M{\"u}nchen, Theresienstr.\ 37, 80333 Munich, Germany}
\newcommand{\harvard}{Department of Physics, Harvard University, Cambridge MA 02138, USA}

\newcommand{\Title} {Scaling of the thermal spectral function for quantum critical bosons in one dimension}
\newcommand{\Authors}
{
\author{Thomas Barthel}
\affiliation{\lmu}
\author{Ulrich Schollw\"{o}ck}
\affiliation{\lmu}
\author{Subir Sachdev}
\affiliation{\harvard}
}
\newcommand{\Date} {December 12, 2012}

\begin{document}

\title{\Title}
\Authors

\date{\Date}

\begin{abstract}
We present an improved scheme for the precise evaluation of finite-temperature response functions of strongly correlated systems in the framework of the time-dependent density matrix renormalization group. The maximum times that we can reach at finite temperatures $T$ are typically increased by a factor of two, when compared against the earlier approaches. This novel scheme, complemented with linear prediction, allows us now to evaluate dynamic correlators for interacting bosons in one dimension. We demonstrate that the considered spectral function in the quantum critical regime with dynamic critical exponent $z=2$ is captured by the universal scaling form $S(k,\omega)={1}/{T}\cdot\Phi_S({k}/{\sqrt{T}},{\omega}/{T})$ and calculate the scaling function precisely.
\end{abstract}


\maketitle

\section{Introduction}
Response functions $\bra \hB(t)\hA\ket$ quantify the effect of distortions $\hA$ of the system at time zero on the expectation values of observables $\hB$ at time $t$. They contain important information on the governing quantum many-body physics \cite{Fetter1971} and are accessible in many different experimental setups. For example, recent advances in neutron-scattering techniques make very precise measurements possible \cite{Zaliznyak2005}. 
It is of high importance to have numerical tools at hand that allow for an efficient and highly accurate computation of response functions for (strongly-correlated) condensed matter models in order to match theoretical models to actual materials and to gain an understanding of the underlying physical processes. Arguably, sufficiently precise tools are available for one-dimensional (1D) systems at zero temperature \cite{Jeckelmann2002-66,Schollwoeck2005,White2008-77,Ren2012-85}. At finite temperatures, which, needless to say, are the relevant case for most experiments, our numerical abilities were however quite limited.
As discussed and demonstrated in several works \cite{Barthel2009-79b,Feiguin2010-81,Karrasch2012-108}, finite-temperature response functions for strongly-correlated 1D systems can be evaluated up to some maximum reachable time by using the time-dependent \emph{density matrix renormalization group} (tDMRG) \cite{Vidal2003-10,White2004,Daley2004}. A difficulty in the simulations of time-evolved states is the growth of entanglement with time \cite{Calabrese2005,Bravyi2006-97,Barthel2008-100}. In tDMRG calculations, this leads to a corresponding severe increase of the computation cost and a strong limitation of the maximum reachable times, depending on the available computational resources. The effect is much more drastic for mixed states: At low temperatures, the corresponding computation cost basically increases by a power of two in comparison to the simulation of pure states. It is decisive to reach times that are sufficiently long to allow for the extraction of the desired physical information like spectral properties.

In this paper, we present a novel tDMRG scheme for the calculation of dynamical correlators at finite temperatures which typically doubles the maximum reachable times in comparison to the earlier schemes in the literature. This allows for a very precise evaluation of thermal response functions which could not be addressed before. We also analyze and explain the computation cost for the different schemes.
As a specific application we study bosons ($[\hb_i^\pdag,\hb_j^\dag]=\delta_{ij}$) with repulsive onsite and nearest-neighbor interactions as described by the extended Bose-Hubbard model
\begin{align}
	\hH=&-\frac{1}{2}\sum_i(\hb^\dag_i\hb^\pdag_{i+1} + h.c.) - \mu\sum_i\hat n_i\nonumber\\\label{eq:H}
	&+ U\sum_i  \hat n_i(\hat n_i-1) + V \sum_i \hat n_i\hat n_{i+1}.
\end{align}
The new tDMRG scheme now allows us to examine the thermal \emph{spectral function} \cite{Fetter1971}
\begin{equation}
	S(x,t)= \frac{1}{2\pi}\Tr\left(\frac{e^{-\beta\hH}}{Z_\beta}\,[\hb_{x+x_0}(t),\hb_{x_0}^\dag]\right),\label{eq:S}
\end{equation}
with the partition function $Z_\beta=\Tr e^{-\beta\hH}$ and the inverse temperature $\beta=1/T$.
\begin{figure}[b]
\includegraphics[width=0.7\linewidth]{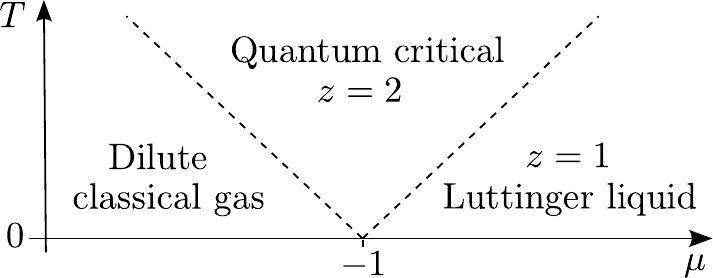}
\caption{\label{fig:phasediag}Crossover phase diagram \cite{Sachdev1999} around the quantum critical point at $\mu=-1$, $T=0$. The $T=0$ state has zero density for $\mu \leq -1$, and non-zero density for $\mu > -1$. In this paper, we compute the universal dynamic correlators in the $z=2$ quantum critical region. Dynamic correlators in the dilute classical gas region were computed in Ref.~\cite{Sachdev1997-78}.}
\end{figure}
The system becomes critical at the point $\mu=-1$, $T=0$ as shown in Fig.~\ref{fig:phasediag}. We will fix $\mu=-1$ in the following. In the $z=2$ quantum-critical region of Fig.~\ref{fig:phasediag}, for small quasi-momenta $k$ and frequencies $\omega$, the temperature $T$ can be expected to be the dominating energy scale and we can expect
\begin{equation}\label{eq:S_Fourier}
	S(k,\omega) := \sum_x e^{ikx} \int_{-\infty}^\infty\ud t\, e^{i\omega t}S(x,t) 
\end{equation}
to be a function of $\frac{k}{\sqrt{T}}$ and $\frac{\omega}{T}$, i.e., $S(k,\omega)\approx f(T)\cdot\Phi_S(\frac{k}{\sqrt{T}},\frac{\omega}{T})$.
As the integral of the spectral function over $\omega$ (at fixed $k$) is unity, we can conclude $f(T)=\frac{1}{T}$.
\begin{equation}\label{eq:S_scale}
	S(k,\omega)\approx\frac{1}{T}\cdot\Phi_S\left(\frac{k}{\sqrt{T}},\frac{\omega}{T}\right)\quad\text{for}\quad k,\omega, T\ll 1
\end{equation}
The existence and possible form of the scaling function $\Phi_S$ has been a long-standing open question.
$\Phi_S$ is expected to be universal for certain (\emph{universality}) \emph{classes} of systems, as the long-ranged correlations at the critical point should not depend on the specific form of the short-ranged interactions. In the language of renormalization group, this means that the renormalization flows of many other systems are governed by the same fixed point.
This allows for a simplification of the calculations by going to the limit $U\to\infty$, for which the system is restricted to have at most one boson per site.
To assert the universality, we have introduced the nearest-neighbor density-density interaction in the model \eqref{eq:H} and will show that the same scaling function applies for several different values of $V$.
It should be stressed that the new tDMRG scheme, described in the following, is generally applicable for finite-temperature real-time and frequency-space simulations for strongly correlated 1D quantum systems.

\section{Method}
In previous tDMRG applications \cite{Barthel2009-79b,Feiguin2010-81,Karrasch2012-108}, the thermal density matrix $\dm_\beta:=e^{-\beta\hH}/Z_\beta$ was encoded by a corresponding \emph{purification} \cite{Uhlmann1976,Uhlmann1986,Nielsen2000,Verstraete2004-6} which is a pure state $|\dmp_\beta\ket\in\H\otimes\H_\aux$ with an auxiliary system $\H_\aux=\H=\Span\{|\vec{\sigma}\ket\}$ such that $\Tr_\aux |\dmp_\beta\ket\bra\dmp_\beta|=\dm_\beta$. A \emph{matrix product state} (MPS) \cite{Accardi1981,Fannes1991,Rommer1997} representation of the purification $|\dmp_\beta\ket$ can be obtained by employing tDMRG for an imaginary time-evolution starting from the infinite temperature state ($\dm_0\propto\id$, $|\dmp_0\ket\propto \sum_{\vec{\sigma}}|\vec{\sigma}\ket\otimes|\vec{\sigma}\ket$) \cite{Verstraete2004-6,Barthel2009-79b,Feiguin2010-81,Karrasch2012-108}. Expectation values then take the form $\Tr\dm_\beta \hat X=\bra\dmp_\beta|\hat X|\dmp_\beta\ket$. The scheme of Ref.~\cite{Barthel2009-79b} to calculate thermal response functions
\begin{equation}\label{eq:chiAB}
	\chi_{\hA\hB}(\beta,t):= \Tr\big(\dm_\beta\,\hB(t)\hA\big) = \Tr\big(\dm_\beta\,e^{i\hH t}\hB e^{-i\hH t}\hA\big)
\end{equation}
consists in obtaining first the MPS purification $|\dmp_\beta\ket$ by tDMRG imaginary-time evolution, and to subsequently do a tDMRG real-time evolution to obtain $|\dmp_\beta,t\ket:=e^{-i\hH t}|\dmp_\beta\ket$ as well as $|A\dmp_\beta,t\ket:=e^{-i\hH t}\hA|\dmp_\beta\ket$. With these, one computes the response function by evaluating the overlap $\chi_{\hA\hB}(\beta,t)=\bra\dmp_\beta,t|\hB|A\dmp_\beta,t\ket$. In the following, we refer to this procedure as \emph{scheme A}.
In this context, it is actually to some extent superfluous to think in terms of purifications \footnote{T. Barthel, \emph{in preparation} (2012).}.\newcounter{foot:inPrep}\setcounter{foot:inPrep}{\value{footnote}} Due to the isomorphism of $\H\otimes\H$ and $\mc{B}(\H)$, the space of linear maps on the Hilbert space $\H$, the MPS occurring in \emph{scheme A} are in one-to-one relation with corresponding \emph{matrix product operators} (MPO), i.e., operators of the form 
\begin{equation}\label{eq:MPO}
	\sum_{\vec{\sigma}\vec{\sigma}'} A^{\sigma_1, \sigma'_1}_1 A^{\sigma_2, \sigma'_2}_2\dotsm A^{\sigma_L, \sigma'_L}_L
	 |\vec{\sigma}\ket\bra\vec{\sigma}'|,
\end{equation}
where the $A^{\sigma_i, \sigma'_i}_i$ are $M_{i-1}\times M_i$ matrices with $M_0=M_L=1$. The sizes $M_i$ are also called \emph{bond dimensions}.
In a short-hand notation, \emph{scheme A} can be denoted by
\begin{equation}\label{eq:schemeA}
	\frac{1}{Z_\beta}\Tr\left(\big[e^{-\beta\hH/2} e^{i\hH t}\big]\hB \big[e^{-i\hH t}\hA e^{-\beta\hH/2}\big]\right).
\end{equation}
The square brackets indicate which parts in this expression are approximated as MPOs [Eq.~\eqref{eq:MPO}] (corresponding to the aforementioned MPS purifications $|\dmp_\beta,t\ket$ and $|A\dmp_\beta,t\ket$) and are obtained via tDMRG. In this notation, the modified \emph{scheme B} used in Ref.~\cite{Karrasch2012-108} reads
\begin{equation}\label{eq:schemeB}
	\frac{1}{Z_\beta}\Tr\left(\big[e^{-\beta\hH/2}\big]\hB \big[e^{-i\hH t}\hA e^{-\beta\hH/2} e^{i\hH t}\big]\right).
\end{equation}
\begin{figure}
\includegraphics[width=1\linewidth]{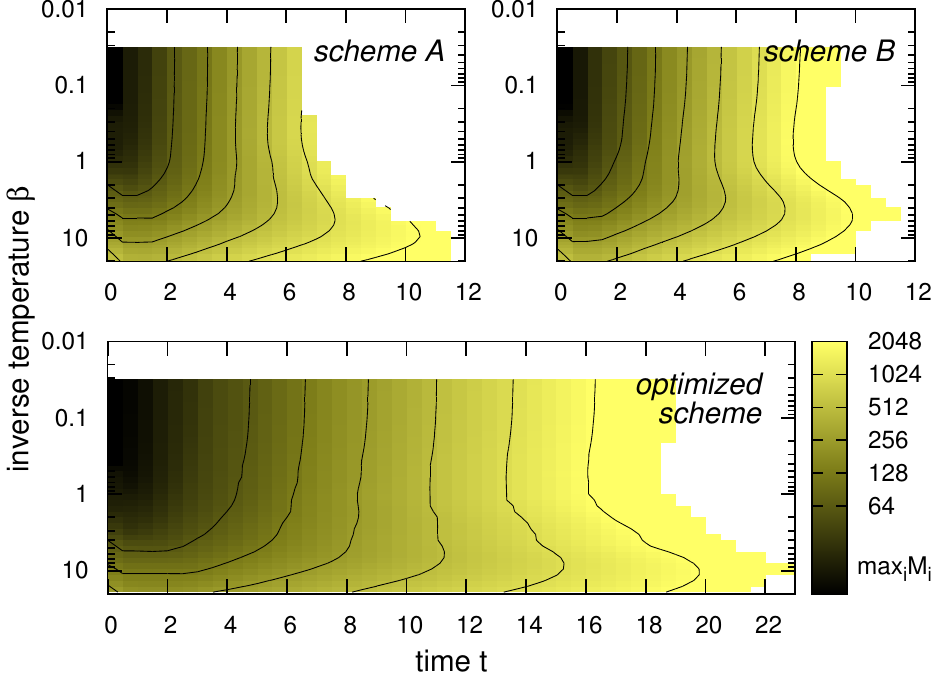}
\caption{\label{fig:dimMax}Maximum bond dimension $\max_iM_i$ occurring in the computation of the response function $\chi_{\hA\hB}$ for $\hB^\dag=\hA=\hb^\dag_{L/2}$, chain length $L=128$, $U\to\infty$, $V=1$, half filling ($\mu=1$), and truncation weights $\epsilon_\beta=10^{-12}$, $\epsilon_t=10^{-10}$. The contour lines mark the values $\max_i M_i=2^6,2^7,2^8,2^9,2^{10},2^{11}$, i.e., the times that can be reached with those maximum bond dimensions. \emph{Schemes A} and \emph{B} reach much shorter times than the scheme of Eq.~\eqref{eq:schemeOpt_param}, optimized with respect to $t'$ and $\beta'$.}
\end{figure}
In each step of the tDMRG, the evolved operators $\hat X = \hat X(t)$ are approximated by an MPO with bond dimensions $M_i=M_i(\beta,t)$ that are as small as possible for a given constraint on the desired precision of the approximation \cite{Schollwoeck2005}. This precision is in each step of the algorithm controlled by the so-called truncation weight $\epsilon=(\normS{\hat X_{\operatorname{trunc}} - \hat X}_2/\normS{\hat X}_2)^2$.
Due to such truncations, the results of the different evaluation schemes like \eqref{eq:schemeA} and \eqref{eq:schemeB} differ slightly from the exact $\chi_{\hA\hB}(\beta,t)$ of Eq.~\eqref{eq:chiAB}. It is essential to keep the errors $\epsilon$ in the MPO truncations controlled and small at all times.

Schemes \eqref{eq:schemeA} and \eqref{eq:schemeB} are typically far from optimal. One has a lot of freedom in designing a scheme that is as efficient as possible. With efficiency we mean that the occurring bond dimensions $M_i$, which determine the computation cost, are as small as possible for given $\beta$, $t$, and $\epsilon$. With a non-singular operator $\hat T$, the most generic splitting involving two MPOs is $\frac{1}{Z_\beta}\Tr\left(\big[e^{i\hH t}\hB\hat T\big] \big[\hat T^{-1}e^{-i\hH t}\hA e^{-\beta\hH}\big]\right)$.
In principle, one would now like to optimize for $\hat T$ in order to minimize the computation cost and, thus, maximize the maximum reachable time. However, in the cases we studied, such optimizations turned out to be inefficient. The required cost scaled exponentially with the system size, even when restricting ourselves to optimize only with respect to unitary operators. Hence, we confine ourselves to study the computation cost of the less general class of schemes
\begin{multline}\label{eq:schemeOpt_param}
	\frac{1}{Z_\beta}\Tr\Big(\big[e^{i\hH t'}e^{-\beta'\hH}\hB e^{-i\hH t'}\big] \\
	\times \big[e^{-i\hH (t-t')}\hA e^{-(\beta-\beta')\hH}e^{i\hH(t-t')}\big]\Big)
\end{multline}
as a function of $t'$ and $\beta'$.
Fig.~\ref{fig:dimMax} compares the evolution of the occurring bond dimensions as a function of temperature and time for the three different schemes, Eqs.~\eqref{eq:schemeA}--\eqref{eq:schemeOpt_param}. In many cases, \emph{scheme B} [Eq.~\eqref{eq:schemeB}] has some advantage over \emph{scheme A}. In Ref.~\cite{Karrasch2012-108} it was pointed out that the involved MPO $\big[e^{-i\hH t}\hA e^{-\beta\hH/2} e^{i\hH t}\big]$ is time-independent for the simple case $\hA=\id$, whereas the computation cost for $\big[e^{-i\hH t}\hA e^{-\beta\hH/2}\big]$, occurring in \emph{scheme A}, can increase with time, even in this trivial case. More generally, the following argument applies for all operators $\hA$ with finite spatial support: Typical condensed matter systems are \emph{quasi-local} \cite{Nachtergaele2007-12a,Barthel2012-108b}, i.e., the spatial support of operators like $e^{-i\hH t}\hA e^{i\hH t}$, occurring in \emph{scheme B}, grows only linearly with time. More precisely, outside a certain space-time cone, the evolved operator acts almost like the identity and does hence not change the entanglement in that region. The preconditions are that all terms in the Hamiltonian are short-ranged and norm-bounded \cite{Nachtergaele2007-12a,Barthel2012-108b}. Nevertheless, as Fig.~\ref{fig:dimMax} indicates, \emph{scheme A} is often advantageous at very low temperatures, especially for non-critical systems. However, the scheme of Eq.~\eqref{eq:schemeOpt_param} optimized for $t'$ and $\beta'$ outperforms the earlier schemes substantially.
A more detailed discussion and analysis of the different evolution schemes will be presented elsewhere \footnotemark[\value{foot:inPrep}]. As the example in Fig.~\ref{fig:dimMax} indicates, the maximum reachable time is a slowly varying function of $\log \beta$ and almost concave. Hence one can reach almost optimal results with \emph{scheme C}
\begin{equation}\label{eq:schemeC}
	\frac{1}{Z_\beta}\Tr\Big(\big[e^{i\hH t_B}e^{-\frac{\beta}{2}\hH}\hB e^{-i\hH t_B}\big] 
	\, \big[e^{-i\hH t_A}\hA e^{-\frac{\beta}{2}\hH}e^{i\hH t_A}\big]\Big),
\end{equation}
which does not require any optimization. After the imaginary time-evolution that yields $[e^{-\frac{\beta}{2}\hH}]$, one runs two real-time tDMRG simulations to obtain MPOs $\big[e^{i\hH t_B}e^{-\frac{\beta}{2}\hH}\hB e^{-i\hH t_B}\big]$ and $\big[e^{-i\hH t_A}\hA e^{-\frac{\beta}{2}\hH}e^{i\hH t_A}\big]$. With Eq.~\eqref{eq:schemeC} one then obtains $\chi_{\hA\hB}(\beta,t_A+t_B)$. The accuracy of the MPOs should be kept under control during the whole simulation, for example, as described above by bounding the truncation error. If this is done properly, it is of minor importance what specific $t_A$ and $t_B=t-t_A$ are chosen to evaluate $\chi_{\hA\hB}(\beta,t)$ for a given time $t$. For the typical case $\hA=\hB^\dag$, the maximum reachable times for $t_A$ and $t_B$ are equal, and the total maximum reachable time with this scheme is then twice as large as the maximum time of \emph{scheme B}. For all simulations in this article we used \emph{scheme C} with a fourth order Suzuki-Trotter decomposition and a time step of size $\Delta t=1/8$ in the tDMRG. The truncation weights were fixed to $\epsilon_\beta=10^{-12}$ in the imaginary time-evolution and $\epsilon_t=10^{-10}$ in the real-time evolution.

\section{Exactly solvable case}
\begin{figure}[t]
\includegraphics[width=0.95\linewidth]{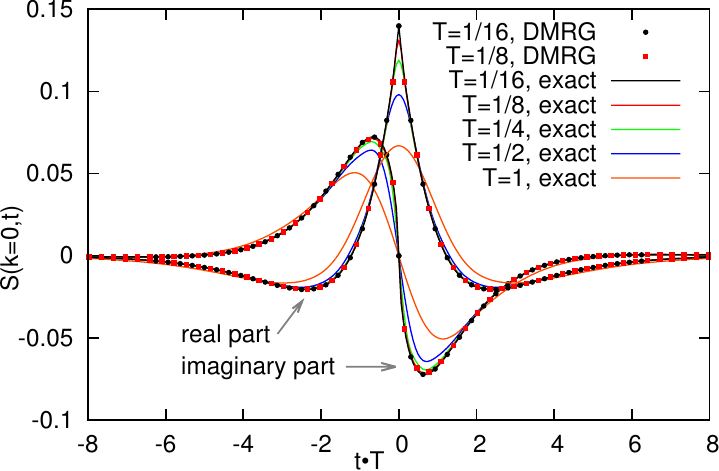}
\caption{\label{fig:Skt_J0_exact}The exact time-dependent spectral function $S(k=0,t)$ plotted as a function of $t\cdot T$ for $\mu=-1$ and $V=0$ and compared with corresponding tDMRG data (\emph{scheme C}).}
\end{figure}
For $U\to\infty$ and $V=0$, the model \eqref{eq:H} can be mapped to a system of free fermions by application of the Jordan-Wigner transformation $\hb_i=\prod_{j=1}^{i-1} (-1)^{\hc_j^\dag\hc_j}\*\,\hc_i$ and the resulting Hamiltonian $-\frac{1}{2}\sum_i(\hc_i^\dag \hc_{i+1}^\pdag+h.c.) - \mu \sum_i \hc_i^\dag \hc_i^\pdag$ can be diagonalized exactly. Due to Wick's theorem, all correlation functions are in this case determined by the single-particle Green's function \cite{Fetter1971}. The response function \eqref{eq:S}, in particular, can be calculated by evaluating Pfaffian determinants of matrices that contain elements of the single-particle Green's function \cite{Caianello1952,Green1964,Stolze1995}. We also use this exactly solvable case to prove the high accuracy of the new generically applicable tDMRG scheme.

\section{Linear prediction}
As the simulations yield the time-dependent spectral function only on a finite time interval $[-t_\max,t_\max]$, defined by the maximum reachable time, a direct Fourier transformation to $S(k,\omega)$, Eq.~\eqref{eq:S_Fourier}, contains \emph{ringing} artifacts. To avoid them, one can use filter functions which, however, result in an artificial broadening. Instead, we use a linear prediction \cite{Yule1927-226,Makhoul1975-63} which basically fits a superposition of damped harmonic oscillations to the data. In the DMRG context, this technique was employed first in Ref.~\cite{White2008-77} for $T=0$ and in Ref.~\cite{Barthel2009-79b} for $T>0$. We then Fourier transform the spectral function after extrapolating it in this way for times $|t|>t_\max$;
\begin{equation}\label{eq:Skw_lp}
	S(k,\omega) = \int_{|t|\leq t_\max}\hspace{-5ex}\ud t\, e^{i\omega t}S(k,t) 
	              \,+ \int_{|t|> t_\max}\hspace{-5ex}\ud t\, e^{i\omega t}S_\LP(k,t),
\end{equation}
where $S_\LP(k,t)$ is the result of the linear prediction. Needless to say, the quality of the linear prediction, and hence the precision of $S(k,\omega)$, depend strongly on $t_\max$.
\begin{figure}[t]
\includegraphics[width=1\linewidth]{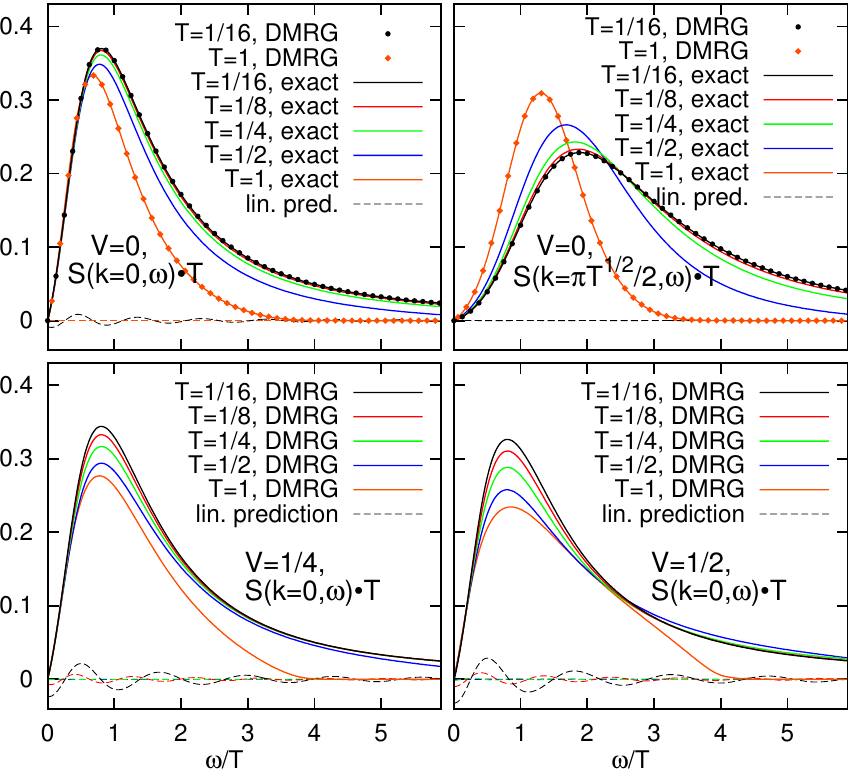}
\caption{\label{fig:Skw}Spectral function \eqref{eq:S_Fourier} for $\mu=-1$ and $V=0,\frac{1}{4}$ and $\frac{1}{2}$, rescaled according to the scaling hypothesis \eqref{eq:S_scale}. The top right plot shows $S(k,\omega)$ for $\frac{k}{\sqrt{T}}=\frac{\pi}{2}$, all others for $k=0$. The dashed lines around $S=0$ show, for the tDMRG data (\emph{scheme C}), the contribution of the linear prediction, i.e., the second term in Eq.~\eqref{eq:Skw_lp}. The simulations were stopped when the computation cost per time step exceeded a certain value. Depending on the temperature, this occurred at maximum bond dimensions $\max_i M_i$ between $1000$ and $4000$.}
\end{figure}

\section{Results for the bosonic spectral function}
In all simulations presented here, we used a system size of $L=128$ and compared against larger systems to ensure that finite-size effects are negligible.
\begin{figure}[t]
\includegraphics[width=0.9\linewidth]{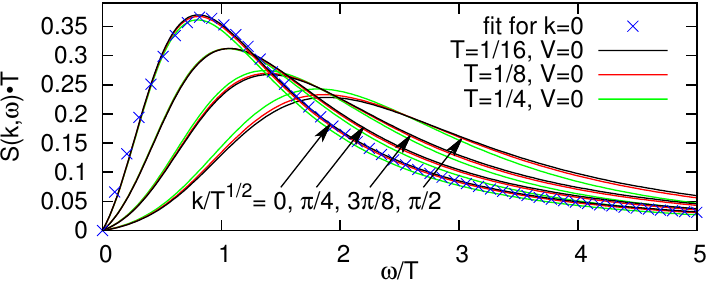}
\caption{\label{fig:Skw_fit}Fitting the scaling function \eqref{eq:S_scale} for $k=0$ with the ansatz $\Phi_S(0,\tilde{\omega}=\frac{\omega}{T})=a\tilde{\omega}/\big(1+b\tilde{\omega}^3\big)$ yields $a=0.649$ and $b=0.826$.}
\end{figure}
Fig.~\ref{fig:Skt_J0_exact} displays the exact solution for $V=0$ and a very clear convergence of the properly rescaled $S(k,t)$ curves. For a lattice size of $L=128$, finite-size effects emerge for temperatures below $T\approx1/32$ as small wiggles at larger times (not shown). The displayed data basically show the behavior in the thermodynamic limit. 
Fig.~\ref{fig:Skw} shows the spectral function in the frequency domain, confirming the scaling hypothesis \eqref{eq:S_scale} by the collapse of the properly rescaled curves for low temperatures, i.e., when plotting $T\cdot S(k,\omega)$ as a function of $\omega/T$ and $k/\sqrt{T}$. The results for $V=\frac{1}{4}$ and $\frac{1}{2}$ confirm the universality of the scaling function $\Phi_S$. The way in which it is approached depends however on $V$. In the limit $T\to 0$ the rescaled curves for different $V$ should coincide.
We would like to stress that in earlier calculations, based on the tDMRG \emph{schemes A} and \emph{B}, we were not able to reach times that would allow for a proper extraction of $\Phi_S$. Especially for low temperatures, the much smaller $t_\max$ in those calculations required a considerably larger contribution of $S_\LP$ to $S(k,\omega)$ [Eq.~\eqref{eq:Skw_lp}] and, furthermore, resulted in relatively big errors of the linear prediction \cite{Yule1927-226,Makhoul1975-63,Barthel2009-79b}. This caused considerable distortions of the curves in comparison to the quasi-exact results displayed in Fig.~\ref{fig:Skw} which were obtained with the novel \emph{scheme C}. Of course, also with the older schemes, one can always reach longer times by increasing the truncation weights $\epsilon_\beta$ and $\epsilon_t$, as this reduces the occurring bond dimensions $M_i$. But for values greater than the ones chosen here, $\epsilon_\beta=10^{-12}$ and $\epsilon_t=10^{-10}$, the precision of the resulting data quickly deteriorates as we have checked by comparison against the exactly solvable case. 
A remarkably good fit of the scaling function $\Phi_S(\frac{k}{\sqrt{T}}=0,\tilde{\omega}=\frac{\omega}{T})$ is given by ansatz $a\tilde{\omega}/(1+b\tilde{\omega}^3)$  with $a=0.649$ and $b=0.826$ as shown in Fig.~\ref{fig:Skw_fit}. For nonzero ${k}/{\sqrt{T}}$, the scaling function differs from this simple ansatz at small $\tilde{\omega}$ but still decays as $\tilde{\omega}^{-2}$ for large $\tilde{\omega}$.

\begin{figure}[t]
\includegraphics[width=1\linewidth]{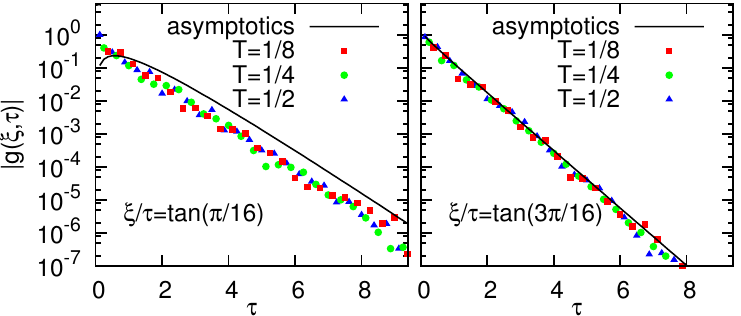}
\caption{\label{fig:Sxt_asympt}Comparison of $T^{-1/2}\bra \hb_{x+x_0}(t)\hb^\dag_{x_0}\ket$ against the analytical formula for the asymptotics as derived by the Riemann-Hilbert problem formalism for $V=0$ \cite{Its1992-54,Korepin1993}.}
\end{figure}
In the continuum limit of the model \eqref{eq:H}, the large time and distance asymptotics of the real-space correlation function $g(\xi,\tau):=T^{-1/2}\bra \hb^\pdag_{x+x_0}(t)\hb^\dag_{x_0}\ket$ with $\xi := x T^{1/2}/2$ and $\tau:=tT/2$ can be evaluated analytically for $U\to\infty$ and $V=0$ in the framework of the Riemann-Hilbert problem \cite{Its1992-54,Korepin1993}. The corresponding formula given in section XVI.9 of Ref.~\cite{Korepin1993} splits into a factor $C_0(\lambda)$ that only depends on the ratio $\lambda:=\xi/2\tau$ and terms that depend on $\lambda$ and $\tau$;
\begin{align*}
	 g(\xi,\tau)
	 =& C_0(\lambda)\tau^{(\frac{1}{\pi}\ln\phi(\lambda)-i)^2/2}e^{2i\tau\lambda^2} \\
	 &\times e^{\frac{1}{\pi}\int_{-\infty}^\infty\ud\mu|2\tau(\mu-\lambda)|\ln\phi(\mu)} \big(1+\mc O(\tau^{-1/2}) \big),
\end{align*}
with $\phi(\mu):=\frac{e^{\mu^2}-1}{e^{\mu^2}+1}$.
The formula for $C_0(\lambda)$ involves a double integral that can not be evaluated easily, but the remaining terms are unproblematic. Fig.~\ref{fig:Sxt_asympt} shows, for space-time lines of constant $\lambda=\xi/2\tau$, a comparison of the analytical formula for the asymptotics of $g(\xi,\tau)/C_0(\lambda)$ against our numerical results for $g$. Both are consistent with each other.

\section{Conclusion}
The presented novel tDMRG scheme \eqref{eq:schemeC} for the evaluation of finite-temperature response functions outperforms earlier approaches substantially and should hence be the method of choice for future applications. In this study, it allowed us to demonstrate that the thermal bosonic spectral function in the quantum critical regime with dynamic critical exponent $z=2$ obeys a universal scaling form and to obtain the corresponding scaling function.

This research was supported by the National Science Foundation under grant DMR-1103860, 
and by U.S. Army Research Office Grant W911NF-12-1-0227.

\bibliographystyle{prsty} 

\end{document}